\documentclass[10pt, twocolumn]{IEEEtran}

\usepackage{lipsum} % For some dummy text
\usepackage{filecontents} % To make the bib-file

\usepackage{times}  
\usepackage{epsfig}
\usepackage{caption}
\usepackage{subcaption}
\usepackage{tabu}
\usepackage{threeparttable}
\usepackage{tabularx}
\usepackage{graphicx} 
\usepackage{color}
\usepackage{xspace}
\usepackage{thumbpdf}
\usepackage{listings}
\usepackage{verbatim}
\usepackage{hyperref}
\usepackage{booktabs}
\usepackage{colortbl}
\usepackage{multirow}
\usepackage{paralist}
\usepackage[colorinlistoftodos, textwidth=3.75cm, shadow]{todonotes}

\hypersetup{pdfstartview=FitH,pdfpagelayout=SinglePage}

\newcommand{\sume}{NetFPGA-SUME\xspace}
\newcommand{\platform}{NetSoC\xspace}
\usepackage{ifthen}

\usepackage{color}

\usepackage{datetime}

\usepackage{xspace}

\newboolean{draft}
\setboolean{draft}{false}

\newboolean{comments}

\ifdraft
  \setboolean{comments}{true}
\else
  \setboolean{comments}{false}
\fi

\ifdraft
\usepackage{calc} 
\usepackage{ifthen} 
\newcounter{hours}\newcounter{minutes} 
\newcommand\printtime{% 
  \setcounter{hours}{\time/60}% 
  \setcounter{minutes}{\time-\value{hours}*60}% 
  \ifthenelse{\value{hours}<10}{0\thehours}{\thehours}:\hspace{-0.33em} 
  \ifthenelse{\value{minutes}<10}{0\theminutes}{\theminutes} 
} 

 \usepackage{draftwatermark}
 \SetWatermarkScale{1.1}
 \SetWatermarkLightness{0.9}
 \SetWatermarkText{\shortstack{DRAFT  \\[0.5cm] {\fontsize{38}{12}\selectfont 
         \today, {\currenttime}}%, \printtime
     }}
  \fi

\begin{document}
\bstctlcite{IEEEexample:BSTcontrol}

%\conferenceinfo{} {}
%\CopyrightYear{}
%\crdata{X}
%\date{}

%%%%%%%%%%%% THIS IS WHERE WE PUT IN THE TITLE AND AUTHORS %%%%%%%%%%%%

%\title{An Exploration of RISC based FPGA Architectures for Networking Applications
\title{Prototyping RISC Based, Reconfigurable Networking Applications in Open Source
}

\author{ 
    \IEEEauthorblockN{Jong Hun Han\IEEEauthorrefmark{1},
                      Noa Zilberman\IEEEauthorrefmark{1},
                      Bjoern A. Zeeb\IEEEauthorrefmark{1},
                      Andreas Fiessler\IEEEauthorrefmark{2},
                      Andrew W. Moore\IEEEauthorrefmark{1}}
                      
    \IEEEauthorblockA{\IEEEauthorrefmark{1}University of Cambridge
    Email: {\emph{firstname.lastname}}@cl.cam.ac.uk}\\
    \IEEEauthorblockA{\IEEEauthorrefmark{2}genua GmbH
    Email: {\emph{andreas}}@leyanda.de}
 } % make the title area

\maketitle

\begin{abstract}

% Whats the problem?
In the last decade we have witnessed a rapid growth in data center systems,
requiring new and highly complex networking devices. 
% Who Cares?
The need to refresh networking
infrastructure whenever new protocols or functions are introduced, and the increasing costs that this entails, are of a concern to all data center providers.
%Whats the solution? 
%\awm{utterly unclear how a new SoC solve the need for periodic refresh}
New generations of Systems on Chip (SoC), integrating microprocessors and
higher bandwidth interfaces, are an emerging solution to this problem. These 
devices permit entirely new systems and architectures that can obviate the replacement of existing networking devices while enabling seamless functionality change. 
In this work, we explore
open source, RISC based, SoC architectures with high performance networking
capabilities. The prototype architectures are implemented on the \sume platform.
Beyond details of the architecture, we also describe the hardware implementation and the porting of operating systems to the platform.
%What criteria will declare success 
%We show %through its evaluation under different test scenarios by
The platform can be exploited for the development of practical networking appliances, and we provide use case examples.

\end{abstract}

\section{Introduction}\label{sec:intro}

Transformational technology~\cite{os-katelin, hp-machine}, combined with computing bottlenecks~\cite{zilberman2016terminating} has led to a
rethinking of the hardware-software designs of modern computing
systems. Undoubtedly, many of these changes had been driven by the growth in networking data
created and consumed by user applications. The increasing amount of networking data also leads architects to focus on data-centric computational architectures able to meet the demand of high I/O performance.
Systems-on-Chip (SoC) integrating high bandwidth network interfaces also emerge on high performance server systems~\cite{costa2015r2c2,xfabric-ms}.

In this paper, we explore architectures embedding RISC based processors on the \sume
platform~\cite{sume-github-web}. RISC processors are dominant in the mobile and portable electronics
appliances markets due to their high power efficiency~\cite{power-emily}. While building a high-performance server based on 64-bit RISC CPUs,
embedded within a SoC alongside high bandwidth integrated networking
and I/O interfaces are also available~\cite{xgen2-am, system-sheng},
%\noa{I think this claim is outdated, and there is SoC+RISC, e.g. X-Gene by applied micro(APM)}.
While the computing performance of x86 based servers is still better,
%\andreas{amd64?ia64?}
it is believed that RISC based servers will eventually prevail thanks to
their superior power efficiency~\cite{shuja2016survey}.

Soft processors provide a high degree of freedom for both the processor design and the architecture at a system level. While free open-source approaches, in contrast to commercial RISC processors (e.g.\ ARM-based), have an appealing low-bar to entry. Extensibility and cost have lead to open-source CPU projects~\cite{riscv-web,lowrisc-web, beri-web} being increasingly attractive to users and developers in both academia and industry. 

The configurability available in commercial processors embedded in FPGAs (e.g. Xilinx's ARM-based Zynq, Altera's Nios) is not sufficient to enable the introduction of new processor-level functionality or for tailoring performance.
In contrast, flexibility can almost verge on excessive with open source processor projects varying significantly: from implementations of an instruction set architecture (ISA), through to a
soft-core processor, to a full SoC implementation. While many of these open
source projects provide tool chains for hardware and software implementations,
a considerable engineering effort is still required to build a system on a newly
customized platform. Even when hardware is already implemented, porting
software to a platform is far from being a trivial task.

Two open source RISC processor implementations are demonstrated on the \sume platform: RISC-V~\cite{waterman2011riscv,lowrisc-web} and BERI~\cite{watson2015beri}. 
%summarized in Table~\ref{tbl:core-comp}.
In each case a significant amount of software operations is executed, including, but not limited to device-drivers, operating-system ports, and a range of application programmes.

A high performance networking platform with an embedded processor adventures to both industry and academic research. As an example, a network switch is a key building block in network systems. In the past, the design of a
switch was focused on the forwarding engine. However, 
network architects are pursuing an increasing level of flexibility without
comprising performance. An embedded processor within the switch enables fully fledge control and adoption of a switch functionality in field, such as the installation of new protocol stacks on the fly.
However, multi-core processor cannot replace the FPGA, as FPGAs provide 
higher data plane performance~\cite{zilberman2015reconfigurable}. 
Although FPGA devices have inherent limitations to achieving high performance
%%computing power 
compared to ASICs, they allow implementation of network fabrics
capable of more than 100\,Gbps (e.g.~\cite{xilinx2014_400G}). Additionally, FPGAs allow verification of line-rate hardware implementations of 
%%may cite xilinx latest announcement for 400Gbps experiment.
system functions without incurring tremendous tape-out costs on state of the art silicon technology.
%
%In this paper, we present a 64-bit MIPS soft core based FPGA architecture for
%network system applications development. The architecture is implemented
%onto a \sume platform. The NetFPGA project~\cite{nf-github-web}
%%which, citation or direct url
%provides software, hardware and community as a basic infrastructure for design,
%simulation and testing, all around an open-source high-speed networking
%platform. The third generation NetFPGA platform, \sume,
%~\cite{zilberman2014sume} %~\cite{zilberman2014sume}
%is a low-cost, PCIe host adapter card able to support 40\,Gbps and 100\,Gbps
%applications.
%
%The soft core processor implemented in this work is an open source core
%designed in Bluespec SystemVerilog (BSV) based on a MIPS R4000, which is known
%as Bluespec Enhanced RISC (BERI)~\cite{watson2015beri}. Our implementation,
%NetBERI, includes peripherals and network fabrics for debug and
%evaluation of networking performance. The processor runs an unmodified FreeBSD
%kernel, which is loaded into a local DDR3 DRAM module.
%

The contribution of this work is several-fold: (i) We demonstrate the development of fully-featured
network-systems, based on an open source RISC based CPU, implemented as an FPGA-based prototype. (ii) We compare and contrast two different open-source processor architectures as the processing cores for the networking systems and (iii) we demonstrate porting and running standard operating systems on top of these fully programmable systems. An integrated open source platform such as this can provide a natural evaluation environment for networking systems as an enabler for their rapid implementation. (iv) We make the infrastructure available as an open source contribution to the NetFPGA ecosystem.

%\noa{saving space, removing non mandatory text} %In section 2, the FPGA architecture with integrated
%modules and the OS are illustrated. Section 3 shows simulation and experimental
%environments and evaluation reulsts. Future works are described in Section 4.

%NetFPGA-SUME platform has various example design for network communication
%applications. Also, a DMA module in the platform, which is one of the main
%cores, is useful to debug, monitor, and control such system on a host PC. We
%show how the NF10 infrastructure is exploited in this work. 

\section{Motivation}

The NetFPGA-SoC (\platform) is a platform for the research community that enables research into current and future approaches to network-based systems. To understand the challenge the platform tackles, consider the following use model:

New network stacks and operating systems for datacenter and cloud computing (e.g.~\cite{belay2014ix,costa2015r2c2,peter2016arrakis}) seek high throughput, low latency and increased application performance. These works take advantage of advances in networking, such as network hardware functionality, flexibility from software defined networks, and novel network fabric topology. However, all these solutions are limited by the architecture imposed by their commodity CPUs. Examples of such limitations include limited support for resource allocation and isolation in the CPU's hardware for I/O or interconnect. 

To study possible solutions to application limitations, a demand is created for platforms that support modifying and extending CPU/systems architectures. This means closed commodity CPU based systems are not suitable. Furthermore, using FPGAs with embedded CPUs makes it hard to port any successful solution to other platforms or adapt it for commercial ASIC use. This calls for an open source CPU based platform.

The CPU is only one part of the equation, as extensive networking knowledge is required to create high performance networking platforms. Processors and networking devices have different performance metrics: IPC vs.\ packets per second, throughput vs.\ bandwidth and latency. The work on \platform brought together researchers from both practices, and through close integration bridged the gap between the fields.

\platform is tailored for scalability studies: if, e.g., there is a $10\times$ frequency ratio between an FPGA and silicon-based CPU, then a design that supports 10GbE on an FPGA will support 100GbE using an ASIC with an embedded CPU. \platform can be used to analysis bottlenecks by reducing only the CPU and data path pipeline frequency without scaling down the DRAM frequency. This changes the relative rates and replicates the impact of improved performance RAM relative to a stable speed of CPU and data-pipeline. In turn, this can assist in detecting hidden architecture constraints, or limitations mistaken to be the memory wall.

In \platform, we support two types of RISC CPUs: BERI and RISC-V (described in section~\ref{sec:arch}). Each of the CPUs supports a different operating system: FreeBSD and Linux, respectively. This achieves three goals: making both the BERI and RISC-V architectures available to networking researchers, making both FreeBSD and Linux available to networking researchers focused on operating systems and network stack design, and enabling comparative performance studies of RISC-V and BERI over the same hardware platform.

\section{System Architecture} \label{sec:arch}

\begin{table}
   \centering
    \caption{Summary of open source RISC processors.}
   \centering
    \label{tbl:core-comp}
   \centering
   \setlength\tabcolsep{2pt} % default value: 6pt
    \begin{tabular}{ >{\centering}p{48px} |  >{\centering}p{105px} | >{\centering\arraybackslash}p{65px} }
    %\begin{tabular}{ l | c | c }
      \hline
                     & BERI~\cite{watson2015beri, mips-jon}  & RISC-V~\cite{waterman2011riscv, riscv-web}  \\
      \hline
      \hline
      ISA            & MIPS R4000                  & RISC-V ISA       \\
      OS             & FreeBSD                     & Linux                 \\
      Design         & Bluespec SystemVerilog (BSV) & Chisel                \\
      FPGA Speed     & 120MHz                      & 50MHz                 \\
      \hline
    \end{tabular}
%\vspace{-1em}
\end{table}

Our networked system is prototyped over \sume~\cite{zilberman2014sume}, %the third generation of NetFPGA platforms,  
%~\cite{zilberman2014sume} is 
a low-cost, PCIe host adapter card able to
support 40\,Gbps and 100\,Gbps applications. 
Two open source RISC processors known as BERI~\cite{watson2015beri} and
RISC-V~\cite{riscv-web}, summarized in Table~\ref{tbl:core-comp}, are
implemented respectively on the \sume platform to explore their performance and
costs. 

\begin{figure*}
\centering
\includegraphics[width=1.6\columnwidth,keepaspectratio]{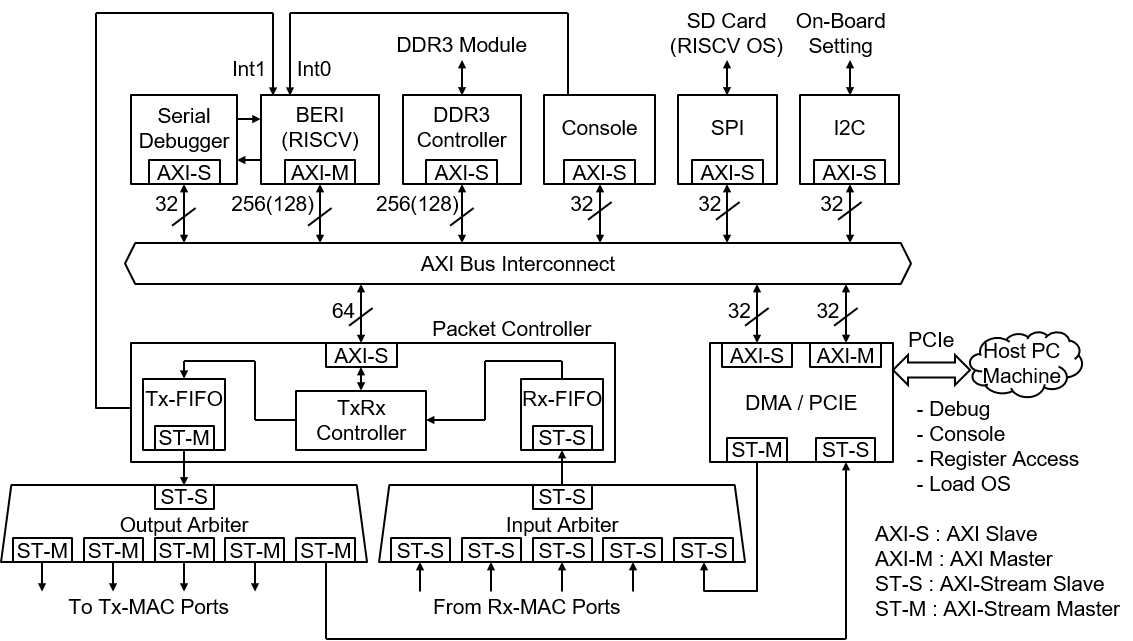}
\caption{RISC processor based FPGA platform architecture.}
\label{fig:sys-arch}
%\vspace{-1em}
\end{figure*}

We use an identical system-level architecture for both RISC processors, as
illustrated in Figure~\ref{fig:sys-arch}. All peripherals have address
mapped registers, exposed to the processors as well as an external host (e.g.
PC) via a PCI-Express (PCIe) interface. In addition to the processors, the
platform integrates multiple NetFPGA modules, such as the networking fabric, as
well and Xilinx peripheral IPs (e.g. DDR memory, 10G port). In the following
subsections, we provide a detailed description of the integrated modules.

%This section illustrates the system architecture for the FPGA implemenation and
%describes each cores incorporated in the system. It consists of basic
%functional blocks for debug, monitor, and input and output interface. A simple
%network interface module with software driver has been developed and integrated
%in the architecture for network communication test. However, we reuse many of
%the existing cores in NF10 and BERI to concentrate system level integration and
%verification. 

%This section illustrates design and functions of our FPGA architecture and
%integrated cores. For the system integration, we reuse several cores in the
%hardware libraries for NetFPGA-SUME reference projects. Direct memory access
%(DMA) core engine incorporating Xilinx Gen2 PCIe endpoing is fully utilized in
%our FPGA implementation. This DMA engine is used in our system for various
%purposes for monitoring, debugging, and controlling the FPGA system as shown in
%Figure 2. It is also used for loading the kernel to the memory and
%communication with serial console driver in the kernel running on the
%processor. The Xilinx MDIO core is to configure Phy chips on the board for the
%ethernet communication, but we have tested the system with default
%configuration.

\subsection{BERI and RISC-V} \label{subsec:beri}

BERI~\cite{watson2015beri} and RISC-V~\cite{riscv-web} are two diverse architectures.
While BERI~\cite{watson2015beri} is designed based on MIPS R4000 Instruction Set Architecture (ISA), RISC-V is
created based on new open source ISA. In~\cite{mips-jon}, BERI is running at a clock speed of 100\,MHz, with a
256-bit wide memory interface and 16\,KB L1 and 64\,KB L2 caches. In this work,
we use the BERI1 flavour of the processor, which is a mature and
higher-performance variant for advanced research~\cite{watson2015beri},
and implement it using 120\,MHz, clock frequency on \sume.
We integrate BERI modules generated by a Bluespec SystemVerilog (BSV) compiler into the
architecture shown in Figure~\ref{fig:sys-arch}.
%The BERI processor is running at a default frequency of \noa{XXMHz} and is can
%execute \noa{XXMOPS/FLOPS}.

%\subsection{Memory Interface} \label{subsec:dram}
Unlike BERI, RISC-V is designed in Chisel~\cite{chisel-web}, which is
an open source hardware construction language that supports layered domain-specific hardware language. A RISC-V ASIC implementation runs at a clock speed of 1.3\,GHz~\cite{lee-riscv}.  In
this paper, we adopted the RISC-V Verilog core variant used in~\cite{lowrisc-web}. In Figure~\ref{fig:sys-arch}, the SPI controller used to access the SD card
storing a bootloader and a Linux kernel is used only by the RISC-V processor.

As a stand alone computing unit, \platform
requires a large memory, from which the operating system and applications can be executed. To this end, we connect the processors to the external
DDR3 SODIMM module running at 1866\,MT/s. The density of each memory module is 4\,GB
%Micron DRR3 spec : MT8KTF51264HZ-1G9E5
%\noa{8GB? for 2 DIMMs?} Jong: We used one of two DDR3 modules in this implementation.
and it can easily accommodate the small size of operating system we
use. The external memory on the platform can be extended up to 32\,GB and used also for other purpose.
%\andreas{IMHO not on SUME (max 2GB per channel)}\noa{no}.
%The memory controller used in the project is based on the DDR3 memory
%controller generated by Xilinx Memory Interface Generator (MIG).
%\noa{version}).

\subsection{SoC Interconnect} \label{subsec:bus}
% AMAM AXI is quite common so it may not need any cite. We also already have many references.
An AMBA AXI protocol is used across the design both for the
~\textit{data-plane} (using an AXI-Stream protocol) and the
~\textit{control-plane} (using AXI4 and AXI4-lite protocols). The AXI bus
interconnect in Figure~\ref{fig:sys-arch} has two master interfaces\footnote
{A RISC-V core has two master interfaces separated into AXI4 and AXI4-lite for memory and peripheral accesses, respectively. One is omitted in Figure~\ref{fig:sys-arch}.}: the
processor(s) and the direct memory access (DMA) engine. The DMA engine is the communication module
with the host machine, using a PCIe interface. Consequently, all the modules
connected to the AXI interconnect can be accessed both by the processor and the
host machine. Therefore, the memory and peripherals can be monitored and
debugged on the host machine side. 

%This interface design is a trade-off between speed and accessibility: while a
%direct interface between the processor and the memory controller would have
%given better processor performance, avoiding transaction delay, the shared
%interconnect enables loading the kernel directly by the host (or from an
%external memory), without wasting processing cycles.  Another advantage is the
%ability to read the memory contents by the host for debug purposes. Similar
%considerations can apply to other slave interfaces connected to the
%interconnect.

The AXI-stream protocol is used for Ethernet packet data transactions across
the architecture. Point-to-point transactions between AXI-stream master and
slave (ST-M and ST-S, respectively) allow to easily handle bursts of data
without compromising the line rate.

%The BERI core was originally designed on Altera DE4 board\noa{cite}, and
%therefore has a master system on chip bus compliant with an Avalon
%protocol\noa{cite} provided by Altera. The Xilinx Virtex~7 FPGA used on \sume
%provides an AXI bus architecture as an on-chip bus. To integrate the two buses,
%we have designed a bus-bridge that converts Avalon into AXI4-lite and vice
%versa. While the physical interface between BERI and the AXI-Interconnect is
%256-bit wide, 256-bit transactions are used only for DRAM memory access.
%Accessing any of the other peripherals is limited to either 32-bit or 64-bit
%(peripheral dependant). This limitation stems from the original BERI design.
%\noa{not sure what of the following is correct, did not touch this text} While
%the processor has 256-bit wide interface to access the memory, data width of
%AXI for Virtex~7 is limited to 32bits and AXI does not support burst data
%transaction. Hence, we have designed an AXI interface controller for the bridge
%to 256-bit wide transaction. However, this leads to degrading performance of
%the processor due to long cycle for accessing the memory. \andreas{AXI4 is not
%limited to 32 bit (lite: 64, full: 1024)}

\subsection{Networking Modules}
The networking modules used in the architecture are different from the
reference modules provided by \sume. While there is some similarity in roles
and functionality, the RISC embedded architecture requires a different
implementation, as illustrated in Figure~\ref{fig:sys-arch}. The networking
modules consist of four module types: 10\,GbE ports, Input Arbiter (IAR), Output
Arbiter (OAR) and Packet Controller (PAC). Unlike a PC handling high bandwidth
networking through PCIe a board, SoC can enable tightly binding CPU, MMU, and
networking interface to improve performance in terms of data rate and latency. 

The 10\,GbE port includes basic Ethernet layers one and two functionality. 
%,including the physical coding sublayer (PCS), physical medium attachment (PMA) and media access control (MAC). 
Every 10\,GbE port module (omitted from
Figure~\ref{fig:sys-arch}), is a combination of an incoming port and an
outgoing port, which create a single physical port. Meta data indicating packet
length and source port information is appended to each packet within this
module as well.
%\andreas{what for?}.

IAR module arbitrates in a round robin manner between all the
10\,GbE input ports to the device, whereas OAR arbitrates between
all output ports while sending outgoing packets. Both IAR and
OAR use an AXI-Stream interface to connect to the 10\,GbE port
through the MAC core.
 
PAC serves two roles: first, it converts all AXI-Stream
transactions from the network to AXI-Lite transactions towards the AXI
interconnect, and second it controls all the transactions from the network to
RISC and the other way around. Any incoming packet arriving from the input
arbiter triggers an interrupt to the CPU, which in turn reads the packet from the
PAC (through the AXI interconnect). Outgoing packets, from
the processor to the network, also pass through the PAC to the
Output Arbiter, an from there to the 10\,GbE ports.
 
In \platform, no packet processing is done by the networking modules, as this functionality is executed by the kernel driver running on the processor.
This feature enables high programmability in the \platform platform, as any protocol or functionality can be programmed into the processor per use case. 

%Figure FIXME shows a block diagram of the ethernet packet interface comprising a
%data control unit, receiver (RX) and transmitter (TX) units for incomming and
%outgoing packet data. The processor can access the ethernet packet interface to
%read and write packet data and meta data following AXI-stream protocol. For
%packet transmission, the processor writes at first the meta data (U) which
%contains packet length and destination port to be forwarded. Then, it writes
%strobe signals (S) indicating byte enable of data and packet data (D0 or D1).
%In this work, each of data has 32bits wide and the strobe signals correspond to
%the width of the data. These writing sequences for strobe signals and packet
%data are repeasted until end of packet. A signal of the last data triggers Tx
%to start sending a packet to output arbiter forwarding to a destination port.
%The reading of the received packet data follows the same process as the
%writing, but in the reverse order. The arrived packet from the input arbiteris
%accumulated in a FIFO. The processor reads meta data first to identify the
%packet size and the source ports and then repeats process of reading the strobe
%signals and data until the last signal is asserted. 

%\noa{I propose to unify the next 2 subsections in order to save space (for results/figures)}
 
\begin{figure}
\centering
\includegraphics[width=0.7\linewidth]{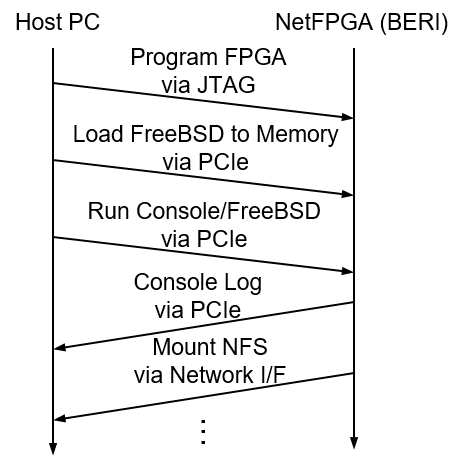}
%\caption{} \label{fig:run-seq} % \hfill%
\caption{BERI FPGA programming and boot sequences.}\label{fig:run-seq}
%\vspace{-1em}
\end{figure}

\subsection{Debugger and Console Modules}

%Two peripherals for debugging and console terminal are designed based on FIFO
%memories. They are dedicate modules for BERI processor. 
BERI processor
operations can be traced and debugged using a built-in debugging
unit~\cite{watson2015beri}. The debugging unit traces the processor operations and provides
access to local registers, interfacing with the serial debugger in
Figure~\ref{fig:sys-arch} by sending and receiving serial byte data. The
serial debugger, connected to the AXI interconnect, is driven by a host PC
through the PCIe interface. Using the serial debugger, the processor can be
paused, resumed and transfer a dump of the operated instructions for debug
purposes. It is also used to trigger the boot process of the BERI core after
loading the FreeBSD kernel into the DDR3 memory.
%\andreas{fixed last par, please check if correct}
%An additional mean of debug is provided through the \noa{NetBeri?} debugging
%unit\noa{is this the "serial debugger" in Fig 1?}. The debugger traces both
%transactions to the memory and AXI bus transactions of the processor, and can
%send them to the host through the PCIe interface, where they can be analyzed.
%This additional debug module is not mandatory for the platform's operation and
%can be eliminated if required by resources' constraints.
 
%\subsection{Console Module}
A console module is designed to emulate a serial interface
peripheral (e.g UART). It is mapped onto the same address range as
the BERI serial interface implemented in~\cite{freebsd-brooks}. The
console module can be used as a terminal of the BERI operating system.
%Like 
%other peripherals in the NetBERI architecture, this serial communication for
%the BERI console also uses the PCIe interface\noa{removed to save space}.

\subsection{Operating System and Kernel Network Driver}

Both processor implementations are running standard operating systems: FreeBSD for BERI, and Linux for RISC-V. %They support a number of processors including MIPS.

While Linux for the RISC-V is used ``as is'' from~\cite{riscv-web}, with FreeBSD for BERI the process was different. We started with the BERI
infrastructure described in~\cite{freebsd-brooks}. Our porting process
followed the adjustments common to bring-up of a new embedded or SoC target
device using an already supported CPU. We had to update the device tree source,
describing the hardware available, and the boot loader, for the memory layout
and the peripherals. A new kernel configuration file was added including
BERI specific peripherals and setting platform specific options. In addition to
the aforementioned console modules, a network interface driver was written for
interconnecting with the platform, which is already publicly available in~\cite{nf-mac-freebsd}. %\url{https://www.freebsd.org/cgi/man.cgi?query=netfpga10g_nf10bmac&manpath=FreeBSD+10.1-RELEASE}.
One advantage of the FreeBSD operating system, is its existing support of on the fly installation of network protocols~\cite{stewart2016tcp}, which largely advantages networking appliances.
%% https://svnweb.freebsd.org/base?view=revision&revision=264601
% XXX-BZ mention in the man page that this also works with SUME once published.
Without a DMA engine on the BERI side, the kernel network driver implements a
classic register-based programmed I/O (PIO) interface talking to an input and
an output FIFO. The evaluation results in Section~\ref{sec:results} are
obtained using this driver.
%\andreas{no ref?}

%Although operating systems software is not within the scope of this paper, it is
%worth mentioning that we ported FreeBSD kernel for the BERI architecture on
%the platform. We port unmodified BERI FreeBSD kernel on the NetFPGA platform
%as describe in ~\cite{freebsd-brooks}. We used existing boot loader with little
%modification for peripherals newly added. We follow most of the
%procedure described in ~\cite{freebsd-brooks} to build the kernel. A MAC
%networking driver for the hardware module is also built in the kernel, which is
%already found in ~\cite{dummy-cite}\noa{didn't understand the sentence + what is the right cite?}. Without DMA engine in the MAC hardware
%driver, the kernel driver for ethernet packets is implemented using a simple
%PIO \noa{acronym??} based register access method. The evaluation results in section 3 are
%obtained using this driver and the hardware module\noa{which module?}.

%\noa{Need to add here RISC's OS text}

%\subsection{Hardware Testing Environment}

\begin{figure}
% \hfill%[t]{0.9\linewidth}
\centering
\includegraphics[width=0.5\linewidth]{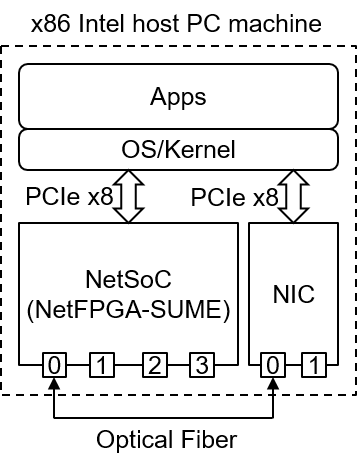}
%\caption{} \label{fig:network-test-setup} 
\caption{Experimental setup for the evaluation.} \label{fig:network-test-setup}
%\vspace{-1em}
\end{figure}

Due to differences in peripherals and drivers available in the OSes for BERI
and RISC-V architectures, their initialization sequences are designed differently. While RISC-V loads an image of the Linux kernel
from the SD card, BERI requires several steps to load its kernel, without using the on-board storage. Figure~\ref{fig:run-seq} illustrates the BERI
initialization sequence during the evaluation (Section~\ref{sec:results}), from programming the FPGA to mounting the
network file system (NFS). The NFS is set between BERI and the host over the 10\,GbE ports shown in
Figure~\ref{fig:network-test-setup}.
%\footnote {A storage device is not yet integrated into the NetBERI architecture}.
%Once the FPGA is programmed, the BERI processor loads an embedded boot loader and
%spins until triggered (e.g. by the debugger) to execute the boot loader
%for the kernel to run.

After the OS kernel is loaded, any application can run on the
target processor, as with any host running FreeBSD or Linux. An immediate example is using the network driver running on FreeBSD to configure and enable the BERI network interfaces. Other examples are using {\it scp} and {\it vi}.
%Running BERI  at 100\,MHz on the platform, it takes about two and a half minutes to boot,
% depending on the FreeBSD kernel features.

%We evaluated the effect of BERI connecting to the memory controller via the AXI
%interconnect versus using direct connection, and found that direct connection is
%about 30\% faster. While this is not negligible, we believe that this is a fair
%trade-off for the additional functionality. 

%\begin{figure} \centering
%\includegraphics[width=0.5\columnwidth,keepaspectratio]{figures/sequence-run-beri.png}
%\caption{BERI programming and boot Sequence.} %\noa{Figure is too small}
%\label{fig:run-seq}
%\vspace{-1em} \end{figure}
%
%\begin{figure} \centering
%\includegraphics[width=0.4\columnwidth,keepaspectratio]{figures/network-experiment-setup.png}
%\caption{Experimental Setup for Network Stack Evaluation.}
%\label{fig:network-test-setup}
%\vspace{-1em}
%\end{figure} %

\section{Evaluation} \label{sec:results}
%\noa{reviewer comment 4: It isn't clear what the clock speed of the BERI processor is.  100 and 120
%MHz are mentioned at various places in the paper.}\\ \noa{fixed?}
We implemented the presented BERI and RISC-V architectures for a Virtex-7 690T
FPGA used on \sume\footnote{This work will be available as a contributed project for the NetFPGA-SUME from the NetFPGA (http://www.netfpga.org) project.}, using Xilinx Vivado 2014.4 EDK, following the conventional
Xilinx tool chain design flow. The BERI and RISC-V based platforms run at
120\,MHz and 50\,MHz, respectively. Although a single-core implementation is
used to port the OS kernels, we also investigate the number of BERI and RISC-V cores that can be simultaneously instantiated on the platform. 
%We further describe the simulation environment
%for the verification and the hardware implementation and evaluation results.

% Figure \noa {which???} shows mapping and place and route results of the
%FPGA. As can be seen Table, the processor is the largest area of 30
%percent.

\subsection{Simulation Environment}

\begin{figure} \centering
\includegraphics[width=0.85\columnwidth,keepaspectratio]{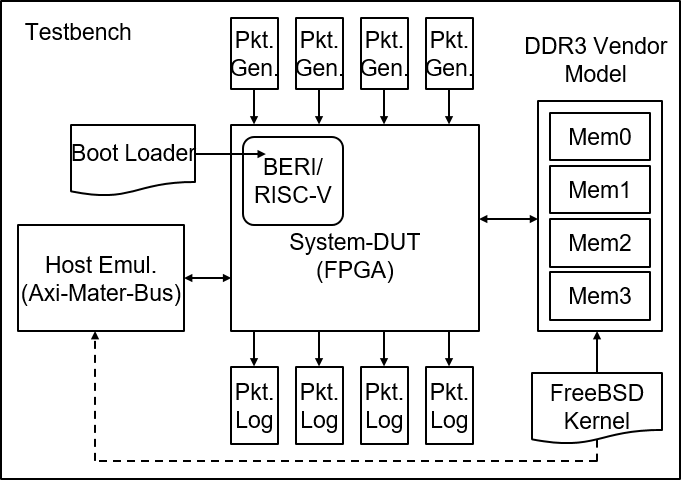}
\caption{RTL level simulation environment for verification.} \label{fig:sim_setup}
%\vspace{-1em}
\end{figure}

Figure~\ref{fig:sim_setup} presents the simulation environment used to
verify the architecture at system level. We conducted a RTL level
simulation, using the same 32\,MB FreeBSD kernel used
in the booting process of the actual hardware. The Design-Under-Test
(DUT) is the top level module implemented on the FPGA. To simulate
interaction with the external memory, the memory controller is
connected to a DDR3 SODIMM memory model~\cite{ddr3-micron}, which represents the same memory as it is used on the SUME board.

While the DRAM memory is initialized in the hardware by the host via the AXI
bus interface, the initialization is not practical in simulation due
to the duration of the simulation processing time. Instead, the memory
model is initialized with the kernel by using a dedicated testbench task.
The task converts the data to a required format, splits the kernel to files and
distributes it as required. In this manner, we can observe in the
simulation the entire core initialization sequence, from BERI entering
the kernel to logging data out into the console module. The RTL level
simulation was performed using the Mentor Questa Sim 10.4 simulator.

The simulation environment for the RISC-V architecture is similar to the
Figure~\ref{fig:sim_setup}. 
%\noa{the following is unclear. does it mean that simulation does not run to the packets stage?}
%The packet simulation can be run under the RISC-V DUT, but it is not meaningful much at the momemnt.
In this case, we verified the sequence until the boot loader
accesses the SD card via the SPI controller, and the Linux kernel is copied into the memory. The simulation environment is also available as open-source as part of the released project.

%The kernel is first prepared by splitting and distributing
%the binary data into four initialization files for the first four
%bank files.
%Special care has to be taken as the data is organized interleaved
%between these files.
%The verilog simulator uses its own file format to represent undefined
%or high impedance values in addition to zeroes and ones in the
%bank files.
%As a consequence, the initilization files have to be handled
%by the verilog testbench, rather than copying them directly.
%This is done by a task in the testbench, which is executed
%after the memory controller signals that the calibration
%process with regard to the memory model has completed successfully.
%According to our experiments, this happens at a simulation time
%of about 600~us. The initialization of the bank files requires 
%virtually no time in the simulation, so that the memory is 
%immediately available after this process. 

\subsection{Multi-Core Implementation Results}

Many-core implementations can benefit a multitude of applications.
We explore multi-core implementations of both processor architectures, and evaluate their resource usage and scalability. Clock speeds of 120\,MHz
and 50\,MHz are used for the single- and multi-core BERI and RISC-V
implementations, respectively. We find that four (quad) BERI cores and eight
(octa) RISC-V cores can be implemented on the platform at a time.
Table~\ref{tbl:costs-comp} summarizes the implementation cost of BERI and
RISC-V, showing the amount of resources taken by the RISC cores, AXI
Interconnect (Inter), Peripheral and Networking modules (Peri) and the remaining unused
resources (Unused). %\noa{does Peri include the networking too?}%Obviously, the
%number of processors affects most of the increase in overall resources. While
%the quad BERI increases LUT and FF by 66.7\% and 13.6\% respectively, the octa
%RISC-V occupies LUT and FF of 75.3\% and 15.4\% in overall FPGA resources. 
The RISC-V implementation is significantly smaller than BERI; this allows
twice the number of cores on the same platform. Differences in the architecture
of every processor affects the surrounding peripherals and interconnects,
leading to a difference in the number of LUT and FF. In addition,
AXI-interconnect (Inter) is affected by the number of cores due to the increased
number of the cores' master interfaces. 

This experiment provides two important insights: First, when a computing-intensive networking application is required, increasing the number of cores in
order to achieve better compute performance is possible. Second, when a
networked data-intensive application is required, a single-core implementation
has plenty of resources (70\%-80\%) available to implement networking and data
processing modules, increasing the bandwidth performance. 

\begin{table}
   \centering
  \begin{threeparttable}
    \caption{Comparison of FPGA implementation cost.}
   \centering
    \label{tbl:costs-comp}
   \centering
   \setlength\tabcolsep{2pt} % default value: 6pt
    \begin{tabular}{ >{\centering}p{26px} |   >{\centering}p{50px} | >{\centering}p{48px} | >{\centering}p{50px} | >{\centering\arraybackslash}p{48px} }
%    \begin{tabular}{ l | c | c | c | c }
      \hline
                     & Single-BERI LUT/FF         & Quad-BERI LUT/FF   & Single-RISC-V LUT/FF      & Octa-RISC-V LUT/FF            \\
%                     & Single             & Quad               & Single             & Octa               \\
%      \hline
%                     & LUT/FF             & LUT/FF             & LUT/FF             & LUT/FF             \\
      \hline
      \hline
      RISC(s)        & $72.1$K/$29.8$K       & $289$K/$118$K         & $40.8$K/$16.7$K       & $326$K/$133$K         \\
       ($\%$)        & $16.6$\%/$3.4$\%  & $66.7$\%/$13.6$\% & $9.4$\%/$1.9$\%   & $75.3$\%/$15.4$\% \\
      \hline
      Inter         & $16.9$K/$18.7$K       & $21.0$K/$19.0$K       & $6.1$K/$6.2$K         & $16.9$K/$13.4$K       \\
       ($\%$)        & $3.9$\%/$2.1$\%   & $4.9$\%/$21.9$\%  & $1.4$\%/$0.7$\%   & $3.9$\%/$1.6$\%   \\
      \hline
      Peri         & $47.8$K/$47.2$K       & $47.5$K/$47.3$K       & $35.8$K/$35.4$K       & $35.6$K/$35.4$K       \\
       ($\%$)        & $11.0$\%/$5.4$\%  & $10.9$\%/$5.4$\%  & $8.2$\%/$4.1$\%   & $8.2$\%/$4.1$\%   \\
      \hline
      Unused         & $296$K/$770$K         & $75.7$K/$681$K        & $350$K/$808$K         & $54.1$K/$683$K        \\
       ($\%$)        & $68.4$\%/$88.9$\% & $17.4$\%/$78.7$\% & $80.8$\%/$93.2$\% & $12.5$\%/$78.9$\% \\
      \hline
    \end{tabular}
  \end{threeparttable}
%\vspace{-1em}
\end{table}

\subsection{Example: Network Stack Evaluation on BERI FreeBSD}

%intel i7-960 octa cores, 24G memory.

% Removed to squeeze the length of the paper.
%A networking appliance using a standard operating system eases the
%implementation of new functions, otherwise costly or hard to
%implement~\noa{cite a network stack implemetation on FPGA paper?}. We use here
%as a test case the use of standard network stack. 

\begin{figure}[ht]
\centering
%\begin{subfigure}[t]{0.45\linewidth}
%\centering
\includegraphics[width=0.8\linewidth]{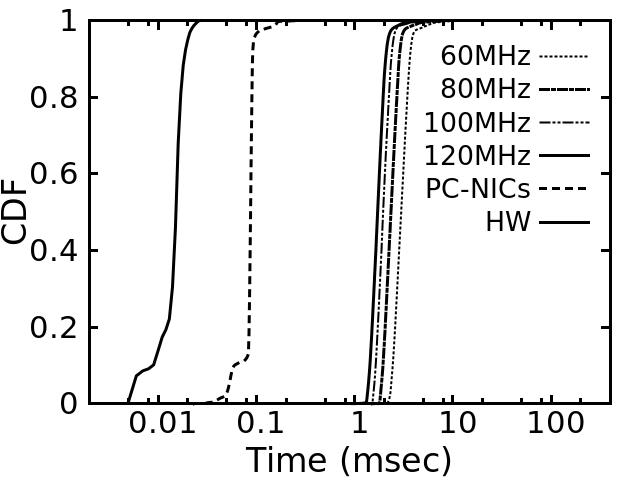}
%\end{subfigure}% \hfill%
%\begin{subfigure}[t]{0.45\linewidth}
%\centering
%\includegraphics[width=\linewidth]{data/ping_120M_latency_result.png}
%\caption{} \label{fig:ping-120M} \end{subfigure} 
%\caption{(a) CDF of ping latency results. (b) CDF and the distribution of ping results using 120MHz BERI core clock.} %\noa{Figure are too small. In B\&W the patterns/lines don't show well }}
\caption{CDF of ping latency result.} \label{fig:ping-total}
%\label{fig:ping-latency} 
%\vspace{-1em}
\end{figure}

The most likely use case of \platform is closely integrating new CPU features with the network. Our experiment therefore uses the BERI processor and network utilities running on FreeBSD, with the goal of showing that \platform performance scales comparably with CPU frequency. Figure~\ref{fig:network-test-setup} shows an experimental setup
for the performance evaluation. The PC machine used for the experiment
ran an octa-core i7-960 CPU, with 24\,GB of RAM, equipped with 
\sume and a dual port SolarFlare 10G NIC card. As illustrated in Figure~\ref{fig:network-test-setup}, \platform and the host PC can
communicate through the 10\,GbE ports over an optical fibre connection. In the
experiment, we evaluated how BERI core clock frequencies affect the
\verb+ping+ packet latency between \platform and PC machine, as shown in Figure
~\ref{fig:ping-total} where ``PC-NICs'' is the latency between identical PC
machines using the identical 10\,GbE NIC cards. ``HW'' refers to a FPGA-based hard-coded implementation of \verb+ICMP echo reply+, implemented by the NAAS-Emu project over \sume\footnote{Included in NetFPGA SUME release 1.4.0}, with a known latency of 1.27us and a jitter of 100ns. The latency marked as ``HW'' is the ping round trip latency between the PC and this implementation. It provides us a reference for the latency caused by the PC, fiber and 10\,GbE ports.
 
\begin{figure}[ht]
\centering
\includegraphics[width=0.8\linewidth]{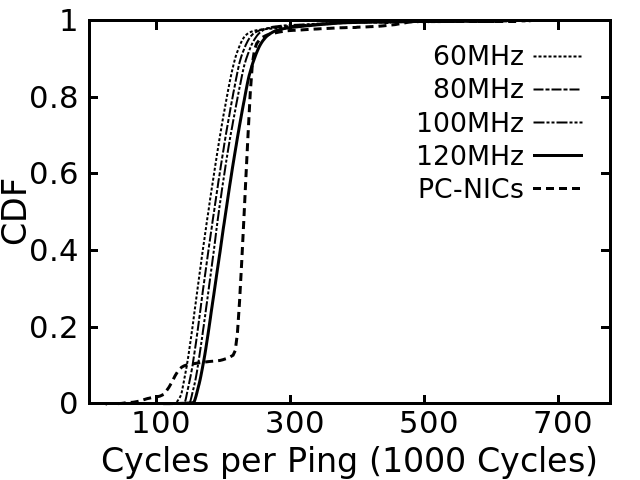}
\caption{CDF of cycles per ping, in thousands of cycles.} \label{fig:ping-cycles}
\vspace{-1em}
\end{figure}

As we consider the scaling of frequency from FPGA to ASIC, we compare the latency of \platform with the latency of PC-NICs in CPU cycles. To improve the accuracy, we only consider the \platform latency, and subtract the median accumulated latency of the requesting PC, fiber and 10\,GbE port latency, represented by ``HW'' (15\,us). The results are presented in Figure~\ref{fig:ping-cycles}: considering the median, there is less than 15\% difference in cycles between \platform default frequency (120MHz) and minimal tested frequency (60MHz), and there is less than 20\% difference in cycles between PC-NICs and \platform running at 120MHz. The higher cycles count of the host is contributed partly to unaccounted contributors, such as the PCIe interconnect.

%Noa:future note: For 100M median is 1.95ms, for 120M median is 1.72ms, 80M median is 2.34ms, 60M median is 2.98ms, PC-PC is 88us, before reducing the 15us.
%so 100M is 193.5K cycles median, 120M is 204.6K, 80M is 186K cycles, 60M median is 177.9K cycles, PC-PC is 255K cycles.  

%
%The minimum latency of "PC-NICs" is
%0.057\,msec, whereas 120\,MHz BERI achieved 1.6\,msec, as shown in
%Figure~\ref{fig:ping-120M}. As standard NICs use ASIC implementation, which %much higher
%clock frequencies, this experiment shows the scaling roadmap for FPGA-based implementation parallel to an equivalent functionality.

%0.057msec
%In our experiments, we evaluated latency of the \platform system using
%different BERI core clock frequencies, as shown in
%Figure~\ref{fig:ping-latency}. \verb+ping+ measurement is used to this end. The
%latencies were measured RTT by \verb+ping+ on the PC machine. As shown in
%Figure ~\ref{fig:ping-total}, the \verb+ping+ latency depends on the BERI core
%clock frequency, and it decreases as the frequency is increased. A minimum
%\verb+ping+ latency of 1.6msec in Figure ~\ref{fig:ping-120M} is measured at
%120MHz core clock frequency.
%A CDF of the \verb+ping+ echo reply latency is
%shown in Figure ~\ref{fig:ping-120M}.
%\noa{what is "PC"? the test is for this line is not clear}.

%\noa{Figure 4b is not clear, esp. the CDF}
We also evaluated TCP and UDP performance on \platform using \verb+iperf+.
The application uses packet processing modules with a simple
register-based PIO and achieves only 6.9\,Mbps for both TCP and UDP.
These results are comparable with FreeBSD running \verb+netsend+ results shown in~\cite{netmap-luigi}. Similarly, we expect to achieve a much higher performance by using userspace applications for packet generation. Beyond those the
achievable data rate can be significantly improved by enhancing the packet
processing module in the networking part of the design, adding features such as
packet DMA to the local RAM, checksum validation, segmentation offload and
more, which are under development for our architecture.

\section{Conclusions and Future work} \label{sec:future}

We presented an open source, RISC based, SoC architectures for networking applications implemented on the \sume
platform. The system was tested using common user applications running on
FreeBSD and Linux operating systems. We showed that the integrated RISC
processor is a feasible solution for networking appliances and that scalable CPU designs can take leverage of \platform. Furthermore, the
flexibility and programmability provided by the platform open new directions
for networking research.

%~\cite{zilberman2016terminating
%\andreas{doubt if fabric is the right term} <- seems ok(jong)
Although the frequency of FPGA devices is limited in comparison with ASICs, their parallel implementation allows evaluating scalable
prototypes and high data rate networking fabrics close to practical systems.
The processors and peripherals presented in this paper require further optimization
through hardware and software co-designing. The ~\textit{data-} and
~\textit{control-plane} designs in the architecture can be considered
relatively independent, and we intend to explore reciprocation within the
integrated architectures. We plan to further extend our work, developing novel
networking fabric solutions as proposed in~\cite{zilberman2016terminating}.
%such as replacing the BERI core with higher performance RISC <- just replacing looks not appeal too much(jong)
%processors~\cite{dummy-cite}.%~\cite{woodruff2014cheri}.

%\noa{future work (to add:)\\
%- Multicore BERI design\\
%- Pkt to DRAM DMA\\
%- Standard NF pipeline (control/data plane for SDN)}

%Double blind revivew does not require the acknowlegement.
%\noa{data needs to become available prior to the CR (if EPSRC funded)}.
% the last sentences may not be suitable for the ending.

\section{Acknowledgements}

We would like to thank the many people who contributed to this paper. We would like to thank Salvator Galea, from the EPSRC NAAS project (EP/K034723/1), who contributed the ICMP echo reply hardware implementation. 
This work was jointly supported by the European Union's Horizon 2020 research and innovation programme 2014-2018 under the SSICLOPS (grant agreement No. 644866), ENDEAVOUR (grant agreement No. 644960), the Leverhulme Trust Early Career Fellowship ECF-2016-289, the Defense Advanced Research Projects Agency (DARPA) and the Air Force Research Laboratory (AFRL), under contract FA8750-11-C-0249. The views, opinions, and/or findings contained in this
article/presentation are those of the author/ presenter and should not be interpreted as representing the official views or policies, either expressed or implied, of the Department of Defense or the U.S. Government.

%\noa{Notes to remember:\\
%\newpage
%- The usual ones: fix/remove all comments, check acronyms, check citations, Oxford comma}
\bibliographystyle{abbrv} 
\bibliographystyle{IEEEtran}
\begin{small}
\bibliography{references}
\end{small}
\label{last-page}

\end{document}